\documentclass[aps,prl,twocolumn,superscriptaddress,floatfix]{revtex4-2}

\usepackage{amsfonts}
\usepackage{amssymb,amstext,amsmath}
\usepackage{graphicx}
\usepackage{bm}
\usepackage{color}
\usepackage{latexsym}
\usepackage{mathtools}
\usepackage{dsfont}
\usepackage[colorlinks=true, allcolors=blue]{hyperref}
\usepackage{braket}

\begin{document}

\title{Interpreting Liouvillian Eigenmodes: Quantized Orbits in Dissipative Systems}

\newcommand{\mpipks}{Max-Planck-Institut f\"ur Physik Komplexer Systeme, 01187 Dresden, Germany}

\author{Ashlin V Thomas}
\affiliation{School of Physical Sciences, National Institute of Science Education and Research, Jatni 752050, India}
\affiliation{Homi Bhabha National Institute, Training School Complex, Anushaktinagar, Mumbai 400094, India}

\author{Felix Fritzsch}
\affiliation{\mpipks}

\author{Masudul Haque}
\affiliation{Institut f\"ur Theoretische Physik, Technische Universit\"at Dresden, 01062 Dresden, Germany}
\affiliation{\mpipks}

\author{Shovan Dutta}
\email{shovan.dutta@rri.res.in}
\affiliation{Raman Research Institute, Bangalore 560080, India}

\begin{abstract}
In closed quantum dynamics, energy eigenstates represent physical excitations which, for systems with a classical limit, localize on quantized orbits in phase space. Such a picture is lacking in open systems: Liouvillians have distinct right and left eigenstates that decay with time and are difficult to interpret. We recover a physical correspondence by introducing a gauge-invariant quasiprobability measure combining right and left eigenstates that interprets a Liouvillian mode as a set of coherences, allowing one to visualize the modes as quasiprobability distributions in a ``doubled'' phase space. Using this measure we show that, remarkably, a damped oscillator retains its quantized ``orbits'' for a large class of linear and nonlinear loss, with measurable signatures. A thermal bath broadens these orbits, similar to energy levels, leading to a characteristic multistage relaxation. This provides an experimentally accessible open-system analog of the eigenstate-orbit correspondence underlying semiclassical quantization. For quadratic systems, our results yield dissipative analogs of invariant tori---a central concept in Hamiltonian systems.
\end{abstract}

\maketitle


\textit{Introduction}|In closed quantum systems, the physical nature of the excited states provides important insights into the dynamical properties \cite{Wolfle2018, Rigol2008, DAlessio2016, Nandkishore2015, Turner2018}. In particular, for integrable systems with a classical limit, semiclassical quantization \cite{Keller1985, Percival1977, berry1972semiclassical} shows that these Hamiltonian eigenstates live on classical energy contours that enclose a quantized area in phase space. Such a picture was crucial for understanding classical-quantum correspondence in the early days of quantum mechanics. In contrast, open quantum systems|which have garnered widespread recent attention \cite{Harrington2022, Szakowski2023, Shovan2025, Daley2014, Sieberer2016, Weimer2021, Fazio2025, Carusotto2025, Stefanini2025, deVega2017}|are described by mixed states (density matrices) that evolve under a Liouvillian, which is generally not Hermitian. As a result, there are distinct right and left eigenstates for a given complex eigenvalue, which do not have any standard physical interpretation. Furthermore, even for the simplest systems with a classical limit (e.g., a damped oscillator) the trajectories in phase space are not closed, so one might expect that the concept of quantized areas breaks down altogether for arbitrarily small damping.

\begin{figure}[t]
    \includegraphics[width=1\columnwidth]{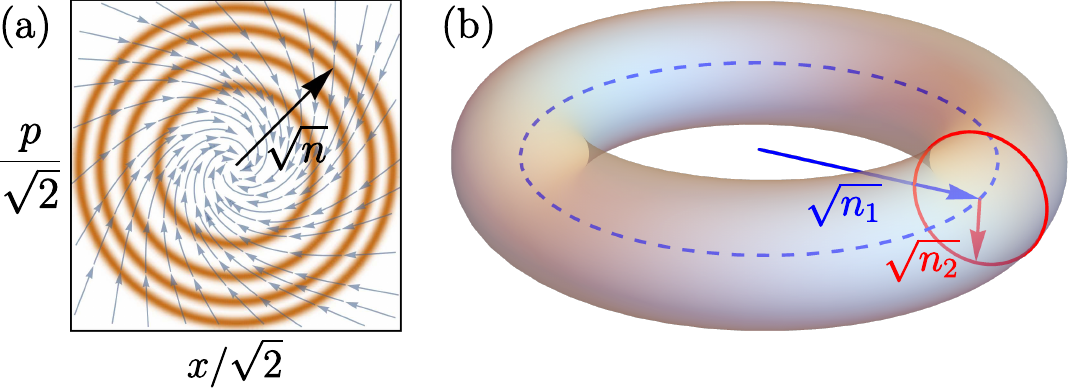}
    \caption{
    (a) Quantized ``orbits'' of a damped oscillator in the backdrop of classical trajectories. A Liouville eigenmode represents coherence between two such orbits or the population of an orbit. (b) For a quadratic system of $N$ oscillators, an appropriately defined probability distribution localizes on a $2N$-torus corresponding to the $2N$ normal modes (here $N=1$).
    }
    \label{Fig1}
\end{figure}

Here we show that, in fact, a physical correspondence exists. First, we introduce a quasiprobability measure associated with a Liouvillian eigenmode, applicable to any open quantum system, which reduces to the standard probability measure in the unitary limit. This allows one to interpret an eigenmode as a distribution of coherences, which can be directly visualized in a ``doubled'' phase space and governs the return amplitude of generic initial states. Second, we show that for an oscillator, area quantization persists for arbitrary nonlinear damping as long as the dynamics is isotropic in phase space [Fig.~\ref{Fig1}(a)]. In this case, a general mode represents coherence between two Fock states (circular ``orbits'') decaying at a characteristic rate. Moreover, the orbits get thermally broadened, similar to energy levels \cite{Demtroder2014}, leading to a multistage, nonexponential relaxation. These predictions are accessible in existing photonic platforms \cite{Brune2008, Wang2008, Delglise2008, Lvovsky2009, Yang2025, Schuster2007, Gely2019}. Third, for quadratic systems of coupled, driven-damped oscillators, our approach leads to a formulation of area quantization for each of the normal modes (invariant tori) [Fig.~\ref{Fig1}(b)], which is a key concept of Hamiltonian semiclassics \cite{Keller1985}.

\textit{Setup}|Although our measures are applicable to any completely positive trace-preserving (CPTP) map, for direct comparisons with the Hamiltonian case we assume a continuous-time Markovian dynamics of the density matrix, ${\rm d}\hat{\rho}/{\rm d}t = \mathcal{L} \hat{\rho}$, where the CPTP condition requires the Liouvillian $\mathcal{L}$ to have the Lindblad form \cite{gorini1976completely, lindblad1976generators}
\begin{equation}
    \mathcal{L}\hat{\rho} = -{\rm i} [\hat{H}, \hat{\rho}] 
    + \sum\nolimits_{\mu} \hat{J}_{\mu} \hat{\rho} \hat{J}_{\mu}^{\dagger} 
    - \frac{1}{2} \big(\hat{J}_{\mu}^{\dagger} \hat{J}_{\mu} \hat{\rho} + \hat{\rho} \hat{J}_{\mu}^{\dagger} \hat{J}_{\mu} \big) \;.
    \label{eq:lindblad}
\end{equation}
Here $\hat{H}$ is the Hamiltonian and $\hat{J}_{\mu}$ are loss processes. For example, a lossy cavity may be described by $\smash{\hat{J}} = \sqrt{\gamma} \, \hat{a}$, where $\hat{a}$ is the photon loss operator and $\gamma$ is the loss rate. This framework is widely used in quantum optics \cite{Carmichael1993, Carmichael1999, gardiner2004quantum, breuer2002theory, Wiseman2009} and, more recently, in many-body physics \cite{Daley2014, Sieberer2016, Weimer2021, Fazio2025, Carusotto2025, Stefanini2025}. 

One can think of $\mathcal{L}$ as a linear (super)operator acting on $\hat{\rho}$. We consider generic cases where $\mathcal{L}$ is diagonalizable. For each eigenvalue $\Lambda_p$, it has a right eigenstate $\smash{\hat{R}_p}$ and a left eigenstate $\smash{\hat{L}_p}$ which are biorthogonal with respect to the Hilbert-Schmidt inner product and normalized as $\text{Tr}(\hat{L}_p^{\dagger} \hat{R}_q) = \delta_{p,q}$. Expanding an arbitrary initial state $\hat{\rho}_0$ in terms of $\{ \smash{\hat{R}_p} \}$, one can find its time evolution as $\hat{\rho}(t) = \sum_p \text{Tr}(\hat{L}_p^{\dagger} \hat{\rho}_0) \, e^{\Lambda_p t} \hat{R}_p$. For physical scenarios, the modes are damped with $\text{Re}\, \Lambda_p \leq 0$. Additionally, for trace preservation there is a steady state $\smash{\hat{R}_0}$ with $\Lambda_0 = 0$.

Two fundamental issues plague the physical interpretation of these eigenmodes: First, the trace-preserving nature of $\mathcal{L}$ implies that $\text{Tr}(\mathcal{L} \hat{A}) = 0$ for any operator $\hat{A}$. With $\hat{A} = \hat{R}_p$ this gives $\Lambda_p \text{Tr}(\hat{R}_p) = 0$, which means all the non-steady right eigenstates are traceless and cannot represent physical states on their own. This is also true of the left eigenstates for unital dynamics, i.e., where $\hat{\mathds{1}}$ is a steady state, since $\text{Tr}(\hat{A} \mathcal{L} \hat{R}_0) = \text{Tr}(\hat{A} \mathcal{L}) = 0$. Moreover, in bosonic systems $\{\smash{\hat{L}_p}\}$ are generally not trace class \cite{prosen2010quantization}. Second, one can rescale $\{ \hat{R}_p, \hat{L}_p\} \to \{c_p \hat{R}_p, \hat{L}_p / c_p^*\}$ with arbitrary $c_p \neq 0$ without affecting the biorthonormality. A physical quasiprobability measure should be invariant under such ``gauge'' transformations \cite{Richter2025}.

\textit{The coherence measure}|Before diving into the dissipative case, it is instructive to look at the eigenstates in the unitary limit, where $\mathcal{L} = -{\rm i}[\hat{H}, \cdot]$. Here, the right and left eigenstates are identical (up to the gauge freedom), with $\hat{R}_{m,n} = \hat{L}_{m,n} = | E_m \rangle \langle E_n |$ and $\Lambda_{m,n} = {\rm i}(E_n - E_m)$, where $\hat{H} |E_n \rangle = E_n |E_n\rangle$. Thus, for $m \neq n$, the Liouvillian eigenmodes represent coherence between two energy levels, which indeed oscillates as $e^{\Lambda_{m,n}t}$. For $m=n$ they describe the conserved populations of these levels. With dissipation, however, the right and left eigenstates are generally unequal and not rank-$1$. Furthermore, the population sector is massively degenerate in the unitary limit, so adding dissipation is a singular perturbation.

To examine whether an interpretation in terms of coherences exists in the dissipative case, it is useful to associate a mixed state to a pure state describing two copies of the system using the correspondence $| i \rangle \langle j| \leftrightarrow |i\rangle \otimes |j\rangle \equiv |i,j)$ (known as vectorization or Choi isomorphism) \cite{Watrous2018, Naves2025}. The two Hilbert spaces can be thought of as the ``ket'' space and ``bra'' space of the original system. Then $\mathcal{L}$ translates to a non-Hermitian Hamiltonian in this doubled space, whose (vectorized) eigenmodes $|\hat{R}_p)$ and $|\hat{L}_p)$ are biorthonormal: $(\smash{\hat{L}_p | \hat{R}_q} ) = \delta_{p,q}$.

A natural gauge-invariant measure for the $p$-th eigenmode is the spectral projector, $\mathcal{F}_p = | \hat{R}_p )( \hat{L}_p |$, in analogy with the probability measure $|E_n \rangle \langle E_n|$ for eigenmodes of $\hat{H}$. From the biorthonormality it follows that $\text{Tr} \, \mathcal{F}_p = 1$. Thus, $\mathcal{F}_p$ defines a joint quasiprobability distribution for the two copies of the system. From completeness of eigenmodes we also have $\smash{\sum_p} \mathcal{F}_p = \mathds{1}$. For a real-valued distribution one can take its Hermitian part, $\mathcal{F}_p^+ = \frac{1}{2}(\mathcal{F}_p + \mathcal{F}_p^{\dagger})$, which has the same properties. Note, in the unitary limit, $\mathcal{F}_{m,n} = \mathcal{F}_{m,n}^+ = | E_m, E_n )( E_m, E_n |$. 

The diagonal elements of $\mathcal{F}_p$ in a tensor-product basis define a ``coherence matrix''
\begin{equation}
    C_p(i,j) \coloneqq ( i,j | \mathcal{F}_p | i,j ) 
    = \langle i | \hat{L}_p | j \rangle^* \langle i | \hat{R}_p | j \rangle \,.
    \label{eq:coherence_matrix}
\end{equation}
As $\mathcal{F}_p$ has unit trace, $C_p(i,j)$ gives the weight of the state $|i,j\rangle$ in the joint distribution. In terms of the original system, it represents the amount of coherence between two basis states $|i\rangle$ and $|j\rangle$, which add up to one, $\sum_{i,j} C_p(i,j) = 1$. Indeed, starting from $\hat{\rho}_0 = |i \rangle\langle j|$, the coherence evolves as $\rho_{i,j}(t) \equiv ( i,j | \hat{\rho}(t) ) =  \sum_p C_p(i,j) e^{\Lambda_p t}$, similar to the return amplitude $\psi_i(t) = \sum_n |\langle i|E_n \rangle|^2 e^{-{\rm i} E_n t}$ for unitary dynamics. From completeness, $\sum_p C_p(i,j) = 1$, so one can also think of $C_p$ as a distribution over the eigenmodes. In the unitary limit, it is simplest in the energy basis, for which $C_{m,n}(i,j) = \delta_{i,m} \delta_{j,n}$.

While $\mathcal{F}_p$ is a superoperator acting on doubled space, its marginals are standard operators that also define useful quasiprobability measures. In particular, tracing over the ``bra'' space gives the marginal distribution of ``kets,'' $\sum_j C_p(i,j) = \langle i | \hat{R}_p \hat{L}_p^{\dagger} | i \rangle$, with the measure $\hat{K}_p = \hat{R}_p \hat{L}_p^{\dagger}$. Similarly, the marginal distribution of ``bras'' is given by $\hat{B}_p = \hat{L}_p^{\dagger} \hat{R}_p$. Both these measures and their Hermitian parts are manifestly gauge invariant and have trace one. In the unitary limit, they select a particular energy eigenstate: $\hat{K}_{m,n} = |E_m \rangle \langle E_m |$ and $\hat{B}_{m,n} = |E_n \rangle \langle E_n |$.

These measures are applicable to any open quantum system, including many-body systems and those without a classical limit. The existence of ``quantized orbits'' relies on additional semiclassical structure, as we discuss below for an oscillator, for which the simple distributions in the unitary limit continue to hold for any isotropic loss. Note that such oscillator modes arise generally in the approach to a semiclassical steady state \cite{Kirton2018, Aspelmeyer2014, Louw2020, Forbes2024}.

\textit{Damped oscillator}|Consider an oscillator mode $\hat{a}$ for which the eigenstates of $\hat{H}$ are the Fock states $|n\rangle$, which represent quantized orbits in phase space. In particular, their Husimi distribution \cite{Husimi1940, gardiner2004quantum}|defined by the overlap with a coherent state $|\alpha\rangle$|is peaked on a circle of radius $|\alpha| \approx \sqrt{n}$, enclosing an area of $2\pi n$.

The simplest and most common form of damping is single-particle loss, described by an operator $\hat{J} = \sqrt{\gamma}\, \hat{a}$ in Eq.~\eqref{eq:lindblad}. The corresponding classical trajectories are found from the equation of motion for $\langle \hat{a} \rangle \equiv (x + {\rm i} p)/\sqrt{2}$ under a mean-field approximation, $\langle \hat{a}^{\dagger m} \hat{a}^n \rangle \approx \langle \hat{a} \rangle^{* m} \langle \hat{a} \rangle^n$ \cite{Drummond1980}. For one-body loss they spiral to the origin [Fig.~\ref{Fig1}(a)] with radial decay $\dot{r} = -(\gamma/2) r$. 

\begin{figure}[t]
    \includegraphics[width=0.97\columnwidth]{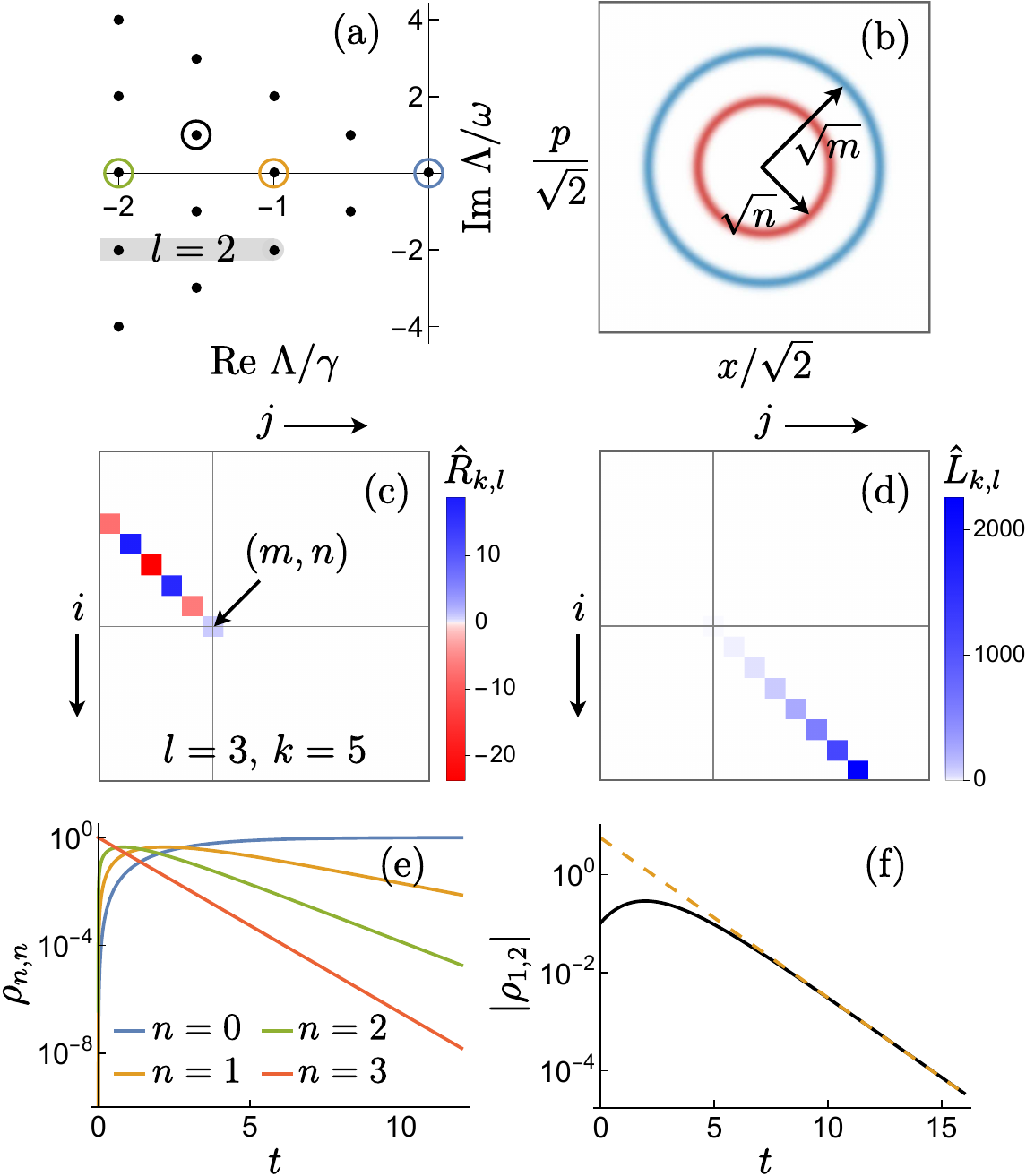}
    \caption{
    (a) Liouville spectrum of a harmonic oscillator with frequency $\omega$ and damping $\gamma$. The imaginary parts are set by the phase winding $l$, corresponding to the $l$-th diagonal in the Fock basis. (b) An eigenmode (for $l \neq 0$) represents coherence between two Fock states $m$ and $n$. (c,d) Corresponding right and left eigenstates overlap only at $(m,n)$. (e) Populations of different Fock states after starting from $|n=3\rangle$ with $\gamma=0.5$, each governed by a specific eigenvalue circled in (a) with the same color. (f) Coherence between Fock states $|1\rangle$ and $|2\rangle$ after starting from the coherent state $|\alpha=2\rangle$ with $\omega=1, \gamma=0.5$. The asymptotic decay (dashed line) is governed by the corresponding eigenmode circled in black in (a).
    }
    \label{Fig2}
\end{figure}

Crucially, this decay is isotropic due to the weak rotational symmetry \cite{Buca2012, Albert2014, Baumgartner2008} $\smash{\hat{U}_{\phi} = e^{{\rm i} \phi \hat{a}^{\dagger} a}}$, which transforms the generators as $\smash{\hat{U}_{\phi}^{\dagger}} \hat{H} \hat{U}_{\phi} = \hat{H}$ and $\smash{\hat{U}_{\phi}^{\dagger}} \hat{J} \hat{U}_{\phi} = e^{{\rm i}\phi} \hat{J}$, leaving $\mathcal{L}$ invariant. As a result, $\mathcal{L}$ cannot mix different sectors of the superoperator $\mathcal{U}_{\phi} \coloneqq \hat{U}_{\phi}^{\dagger} (\cdot) \hat{U}_{\phi}$. The eigenstates of $\mathcal{U}_{\phi}$ are operators $| m \rangle \langle n |$ with eigenvalue $\smash{e^{{\rm i } \phi (n-m)}}$, so a given sector corresponds to operators that live on a particular diagonal, $l = m-n$, in the Fock basis. Hence, each diagonal has their own set of right and left eigenstates.

For a harmonic oscillator with $\hat{H} = \omega \hat{a}^{\dagger} \hat{a}$, these eigenvalues and eigenstates can be solved analytically \cite{zubairy1980photon, Briegel1993, Endo2008, Honda2010}, with $\Lambda_{k,l} = -{\rm i} \omega l - \gamma (k + |l|/2)$. Here $k$ labels the different eigenstates for a given diagonal, $k = 0,1,2,...\,$. Figures \ref{Fig2}(a,c,d) show the spectrum and an eigenstate pair. The wedge-shaped point spectrum is characteristic of systems with a stable fixed point in the classical limit \cite{Dutta2025}.

Importantly, $\hat{R}_{k,l}$ has support over only the first $k+1$ states in the $l$-th diagonal, whereas $\smash{\hat{L}_{k,l}}$ has support from the $(k+1)$-th state up to infinity. This structure follows from the unidirectional flow in Fock space|one can only flow to lower excitations. Hence, $\smash{\hat{R}_{k,l}}$ is an eigenvector of an upper triangular rate matrix and has support in the future ``light cone'' of $|k+l\rangle \langle k|$ (for $l \geq 0$; see Appendix A for more details). Conversely, $\hat{L}_{k,l}$ has support in the past ``light cone''; they are right eigenstates of $\mathcal{L}^{\dagger}$.

The above structure leads to a key simplification: The coherence matrix in Eq.~\eqref{eq:coherence_matrix} involves element-wise overlap of $\hat{R}_{k,l}$ and $\hat{L}_{k,l}$, which is nonzero only for $|m \rangle\langle n|$ where $l=m-n$ and $k = \text{min}(m,n)$. Furthermore, this overlap must be $1$ as the matrix elements add up to unity. Thus, even though the right and left eigenstates are very different from the unitary case, the spectral projector still represents coherence between two specific Fock states for $l \neq 0$ [Fig.~\ref{Fig2}(b)] and population of a Fock state for $l=0$, with a one-to-one mapping between $(k,l)$ and $(m,n)$.

This interpretation has measurable consequences: (1) starting from $\hat{\rho}_0 = |n \rangle\langle n|$ (corresponding to $k=n, l=0$), the survival probability falls as $\rho_{n,n}(t) = e^{\Lambda_{k,l} t} = e^{-n \gamma t}$. This linearly growing decay rate \cite{lu1989effects} has been observed in microwave resonators \cite{Brune2008, Wang2008}. (2) For generic initial states, the late-time relaxation of $\rho_{m,n}(t)$ is governed by the slowest-decaying mode for which $\smash{\hat{R}_{k,l}}$ has support in $|m \rangle\langle n|$; this is also the mode with $C(m,n) = 1$. Hence, $\rho_{m,n}(t) \sim e^{\Lambda_{k,l} t}$ where $(k,l)$ corresponds to $(m,n)$. This is shown in Figs.~\ref{Fig2}(e-f) and may be tested experimentally using Wigner tomography in photonic setups \cite{Delglise2008, Lvovsky2009, Yang2025}.

\textit{Generality}|The selection of a unique pair of Fock states in the coherence matrix relies only on the isotropy and unidirectional flow, which holds for a larger class of nonlinear models. The one-way flow in Fock space occurs for any number-conserving Hamiltonian and loss operators of the form $\hat{J} = \sum_{\mu, \nu = 0}^{\infty} A_{\mu, \nu} \hat{a}^{\dagger \mu} \hat{a}^{\nu}$ where $\hat{A}_{\mu, \nu} = 0$ for $\mu > \nu$. These operators transform under rotation as $\smash{\hat{U}_{\phi}^{\dagger} \hat{a}^{\dagger \mu} \hat{a}^{\nu} \hat{U}_{\phi} = e^{{\rm i}\phi (\nu-\mu)} \hat{a}^{\dagger \mu} \hat{a}^{\nu}}$. Hence, isotropy requires each term in $\hat{J}$ to have the same ``charge'' $q \coloneqq \nu-\mu$. This still includes dephasing ($q=0$) and $q$-body loss ($q > 0$) with arbitrary state-dependent rates.

For instance, take the widely studied Kerr oscillator, $\hat{H} = \omega \hat{a}^{\dagger} \hat{a} + \kappa \hat{a}^{\dagger} \hat{a}^{\dagger} \hat{a} \hat{a}$, with one- and two-body loss: $\hat{J}_1 = \sqrt{\gamma_1} \, \hat{a}, \hat{J}_2 = \sqrt{\gamma_2} \, \hat{a}^2$ \cite{Dodonov1997, Kheruntsyan1999, Bartolo2016, Scarlatella2019, Roberts2020, Tokieda2026}. The classical trajectories are again spirals but with radial decay $\dot{r} = -\frac{\gamma_1}{2} r - \gamma_2 r^3$, which gives power-law relaxation for $\gamma_1 \to 0$. For the quantum problem, the support structure of the eigenstates are exactly the same as before, and they pick the same circular orbits in phase space. However, their decay rates are different: The $k$-th Fock state decays with the eigenvalue $\Lambda_{k,0} = -\gamma_1 k - \gamma_2 k (k-1)$ (see Appendix A for the full spectrum). In line with the classical relaxation, orbits farther from the origin lose population much faster and the decay rates ($|\text{Re}\,\Lambda_{k,l}|$) do not depend on $\omega$ or $\kappa$.

Note that one can apply a Bogoliubov transformation or squeezing \cite{drummond2013quantum} to make the energy contours elliptical. If the loss operators are also transformed, the Liouvillian spectrum is unaltered and the arguments regarding the structure of eigenmodes work the same way in terms of the squeezed Fock states \cite{Kim1989}, which are peaked on the elliptical orbits with the same quantized area.

\textit{Thermal broadening}|The unidirectional flow in Fock space is lost in the presence of a thermal bath with finite inverse temperature $\beta$, which produces a competition of particle injection and loss, described by $\hat{J}_+ = \smash{\sqrt{\gamma_+}} \, \hat{a}^{\dagger}$ and $\hat{J}_- = \sqrt{\gamma_-} \, \hat{a}$ with $\gamma_+ / \gamma_- = e^{-\beta \omega}$ \cite{Carmichael1999}. Classical trajectories continue to spiral inward with the radial decay rate $\gamma = \gamma_- - \gamma_+$. However, the quantum dynamics is modified by the injection. This effect is seen most clearly for a  harmonic oscillator by writing $\mathcal{L}$ as a quadratic form of the ``ket'' and ``bra'' oscillator modes: $a_K \hat{\rho} \coloneqq \hat{a} \hat{\rho}$ and $a_B \hat{\rho} \coloneqq \hat{\rho} \hat{a}^{\dagger}$, which are bosonic superoperators. For $\gamma_+ = 0$, $\mathcal{L} = -{\rm i} \omega (\smash{a_K^{\dagger}} a_K - \smash{a_B^{\dagger}} a_B) + \gamma (a_K a_B - \frac{1}{2} \smash{a_K^{\dagger}} a_K - \frac{1}{2} \smash{a_B^{\dagger}} a_B)$ [see Eq.~\eqref{eq:lindblad}], which is diagonalized as
\begin{equation}
    \mathcal{L} = S \big[ -(\gamma/2 + {\rm i}\omega) \, a_K^{\dagger} a_K - (\gamma/2 - {\rm i}\omega) \, a_B^{\dagger} a_B \big] S^{-1} \;,
    \label{eq:normal_modes}
\end{equation}
with $S = e^{-a_K a_B}$. Thus, $a_K$ and $a_B$ are also the normal (master) modes \cite{prosen2010quantization} up to a similarity transformation, and the eigenstates of $\mathcal{L}$ are characterized by their excitation quanta: $|\hat{R}_{n_K, n_B}) = S |n_K, n_B)$, $(\hat{L}_{n_K, n_B}| = ( n_K, n_B | S^{-1}$, and $\Lambda_{n_K, n_B} = -{\rm i \omega} (n_K - n_B) - \frac{\gamma}{2}(n_K + n_B)$. Indeed, such an eigenmode represents coherence between the Fock states $|n_K\rangle$ and $|n_B\rangle$ of the original oscillator: $C_{n_K, n_B}(i,j) = \delta_{i,n_K} \delta_{j, n_B}$ (see Appendix B). For $\gamma_+ > 0$, $\mathcal{L}$ is still quadratic and diagonalizes as Eq.~\eqref{eq:normal_modes} but with $S = e^{-a_K a_B} \smash{e^{n_T a_K^{\dagger} a_B^{\dagger}}}$, where $n_T$ is the thermal occupation, $n_T \coloneqq 1/(e^{\beta \omega} - 1)$. Due to the isotropy, the right and left eigenstates are again confined to a given diagonal, $l = n_K - n_B$, but they now overlap over a range of states. The resulting distribution has a closed-form expression in terms of Meixner polynomials (see Appendix C). As shown in Fig.~\ref{Fig3}(a), it is positive semidefinite and significant over a window of energy levels $(j_-, j_+)$. For $n \gg l$, the mean and standard deviations are $\overline{j} \approx n / \tanh(\beta \omega  / 2)$ and $\sigma_j \approx n /[\sqrt{2} \sinh(\beta \omega / 2)]$. Both scale as $(\beta \omega)^{-1}$ for a hot bath ($\beta \omega \ll 1$), whereas for a cold bath, $\bar{j} = n + O(\eta)$ and $\sigma_j = O(\sqrt{\eta})$ where $\eta \coloneqq e^{-\beta \omega}$. Hence, the effect of weak thermal fluctuations is to broaden each quantized orbit, similar to spectral broadening of Hamiltonian systems \cite{Demtroder2014}, although the ``lineshapes'' are not Lorentzian. The broadening leads to a multistage, nonexponential decay of a Fock state with signatures of $\overline{j}$ and $j_-$ [Fig.~\ref{Fig3}(b)] (more details in Appendix C), which may be observed in lossy resonators via existing techniques \cite{Brune2008, Wang2008, Schuster2007, Gely2019}.

\begin{figure}[t]
    \includegraphics[width=1\columnwidth]{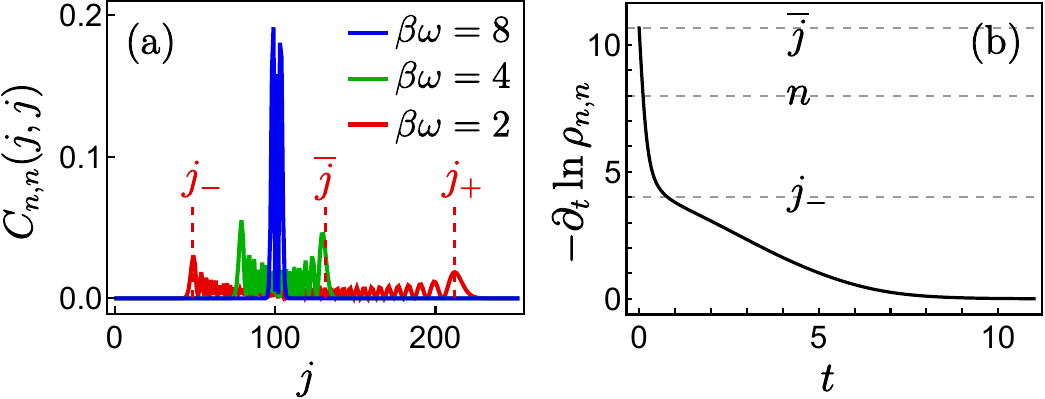}
    \caption{
    (a) Thermal broadening of the eigenmode with $m=n=100$ in the distribution over Fock states $|j\rangle$ at different inverse temperatures $\beta$. (b) Instantaneous decay rate of the return probability of the Fock state $|n=8\rangle$ for $\beta \omega = 2$, $\gamma=1$.
    }
    \label{Fig3}
\end{figure}

\textit{Super Husimi function}|The distribution in the basis of coherent states defines a complex-valued Husimi function, $C_p (\alpha_K, \alpha_B) = ( \alpha_K, \alpha_B | \mathcal{F}_p | \alpha_K, \alpha_B )$, in the doubled phase space. Using properties of coherent states, one can show that (see Appendix D) (i) the marginal distribution of the absolute values $r_K \coloneqq |\alpha_K|$ and $r_B \coloneqq |\alpha_B|$ is given by $A_p(r_K, r_B) = \sum_{i,j} C_p(i,j) Q_i(r_K) Q_j(r_B)$, where $Q_n$ is the Husimi function of a Fock state $|n\rangle$, which is peaked on a circle of radius $r \approx \sqrt{n}$. Hence, the quantized orbits are weighted by the coherence matrix in the Fock basis. (ii) If the dynamics is isotropic, $C_p(\alpha_K, \alpha_B)$ depends only on $r_K$, $r_B$, and the relative phase $\varphi = \arg \alpha_K^* \alpha_B$, allowing three-dimensional visualization. While more general operator bases may be constructed in the doubled phase space \cite{saraceno2016representation}, the coherent-state basis used here directly links $\mathcal{F}_p$ to the quantized orbits.

For a linearly damped harmonic oscillator the radial marginal factorizes, $A_{n_K, n_B\!}(r_K, r_B) = Q_{n_K\!}(r_K) Q_{n_B\!}(r_B)$, and the marginal distributions of $\alpha_K$ and $\alpha_B$ are simply $Q_{n_K}(r_K)$ and $Q_{n_B}(r_B)$. Interestingly, the full distribution|which can be found in closed form|is not peaked at $r_K \approx \smash{\sqrt{n_K}}$ and $r_B \approx \smash{\sqrt{n_B}}$; instead, for $n_K = n_B = n \gg 1$, the peaks are at $r_K^{\star} = r_B^{\star} \approx [n (1+1/\sqrt{2})]^{1/2}$ and $\varphi^{\star} \approx \pm 0.13\pi$ [Fig.~\ref{Fig4}(a)] (see Appendix D for more details). There is no contradiction, however, as $C$ is nonzero everywhere and highly oscillatory at the peaks.

\begin{figure}[t]
    \includegraphics[width=0.9\columnwidth]{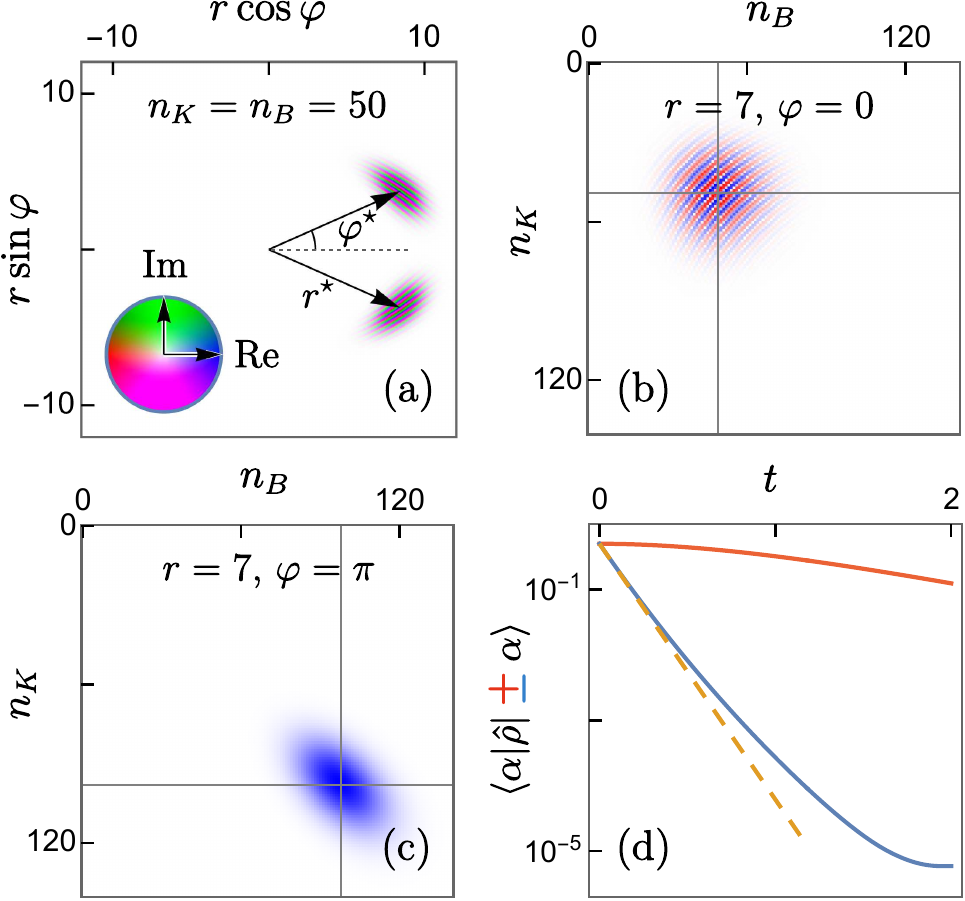}
    \caption{
    (a) A section of the super Husimi function $C_{n_K, n_B}(\alpha_K, \alpha_B)$ in doubled phase space (see text) for the eigenmode that represents population of the Fock state $|n=50\rangle$. Here, $r = |\alpha_K| = |\alpha_B|$ and $\varphi = \text{arg}(\alpha_K^* \alpha_B)$. Arrows show the peak locations at $r^{\star} > \sqrt{n}$ and $\varphi^{\star} \neq 0$. (b,c) Distribution over the eigenmodes for a given $r$ and $\varphi$, peaked at $n_K^{\star} = n_B^{\star} \approx r^2$ for $\varphi = 0$ and $n_K^{\star} = n_B^{\star} \approx 2r^2$ for $\varphi = \pi$ (gray lines). (d) Starting from a cat state $|\alpha\rangle + |\!-\alpha\rangle$ with $\alpha=3$, $\omega=0$, $\gamma=0.5$, the population $\langle \alpha | \hat{\rho} | \alpha \rangle$ decays quadratically, whereas the coherence $\langle \alpha | \hat{\rho} | -\alpha \rangle$ decays at a rate given by the dominant eigenmode for $\varphi = \pi$ (dashed line).
    }
    \label{Fig4}
\end{figure}

The Husimi function connects to the dynamics via the return amplitude $\zeta \coloneqq ( \alpha_K, \alpha_B | \hat{\rho}) = \sum_p \! C_p(\alpha_K, \alpha_B) e^{\Lambda_p t}$. Thus, it is useful to consider the distribution over eigenmodes for a given pair of coherent states. For $\alpha_K = \alpha_B = r e^{i \theta}$, this is indeed peaked at $n_K^{\star} = n_B^{\star} \approx r^2$ [Fig.~\ref{Fig4}(b)]. Conversely, for $\varphi = \pi$, the peak is at $n_K^{\star} \approx r_K^2 + r_K r_B$, $n_B^{\star} \approx r_B^2 + r_K r_B$ [Fig.~\ref{Fig4}(c)]. Additionally, the distribution is positive definite for $\varphi = \pi$, so the dominant eigenmode governs the Loschmidt echo, $|\zeta|^2 \sim e^{-\gamma (r_K + r_B)^2 t}.$ In general, the initial decay rate of $\zeta$ is set by the first moments of $n_K$ and $n_B$, which gives $|\zeta|^2 \sim e^{-\gamma |\alpha_K - \alpha_B|^2 t}$ (see Appendix D). This means, in particular, that starting from a cat state of $|\alpha\rangle$ and $|-\alpha\rangle$, the coherence $\langle \alpha | \hat{\rho} | -\alpha\rangle$ falls as $\smash{e^{-2\gamma |\alpha|^2 t}}$, whereas the populations $\langle \pm \alpha | \hat{\rho} | \pm \alpha\rangle$ decay quadratically, as shown in Fig.~\ref{Fig4}(d).

Analogous doubled-phase-space representations may be constructed for spin systems and even for fermions by extending standard phase-space distributions \cite{Koczor2020, Agarwal1981, Dowling1994, Varilly1989, Tilma2016, Davis2021, Wang2022}.

\textit{Left-right Husimi and invariant tori}|For quadratic $\mathcal{L}$, one can construct a different probability distribution where each of the normal modes localizes on a quantized orbit. As in Eq.~\eqref{eq:normal_modes}, such a Liouvillian can be written in terms of the normal modes $a_1, a_2$ (not necessarily $a_K, a_B$) as $\mathcal{L} = S ( \lambda_1 a_1^{\dagger} a_1 + \lambda_2 a_2^{\dagger} a_2) S^{-1} = \lambda_1 b_1^{\prime} b_1 + \lambda_2 b_2^{\prime} b_2$, where $b_j \coloneqq S a_j S^{-1}$ and $\smash{b_j^{\prime} \coloneqq S a_j^{\dagger} S^{-1}}$ are pseudo-boson modes \cite{Trifonov2009, Bagarello2010, bagarello2017concise, Bagarello2022} that satisfy $[b_i, b_j^{\prime}] = \delta_{i,j}$. They have two sets of coherent states corresponding to the two types of annihilation operators, $b_j$ and $b_j^{\prime \dagger}$: $b_j |\alpha_1, \alpha_2)_R = \alpha_j |\alpha_1, \alpha_2)_R$ and $\smash{b_j^{\prime \dagger} |\alpha_1, \alpha_2)_L = \alpha_j |\alpha_1, \alpha_2)_L}$. These are simply related to the coherent states of $a_j$ as $|\alpha_1, \alpha_2)_R = S |\alpha_1, \alpha_2)$ and ${}_L ( \alpha_1, \alpha_2 | = ( \alpha_1, \alpha_2| S^{-1}$. Hence, the ``left-right'' Husimi function ${}_L ( \alpha_1, \alpha_2 | \mathcal{F}_{n_1, n_2} | \alpha_1, \alpha_2)_R \propto Q_{n_1}(|\alpha_1|) Q_{n_2}(|\alpha_2|)$ localizes on a torus of radii $\approx \sqrt{n_1}$ and $\sqrt{n_2}$ [Fig.~\ref{Fig1}(b)], which is the surface with a given number of excitation quanta of the normal modes, embedded in the ``energy surface'' with $E = \lambda_1 n_1 + \lambda_2 n_2$. This construction readily extends to arbitrary quadratic Liouvillians of $N$ coupled oscillators for which $\mathcal{L} = \sum_{i=1}^{2N} \lambda_i b_i^{\prime} b_i$~\cite{prosen2010quantization}. Here, the left-right Husimi function localizes on a $2N$-torus. This gives a generalization of semiclassical quantization for separable systems \cite{Keller1985} to the dissipative setting and complements time-domain studies of dissipative trajectories in multimode systems \cite{neto2023exact}.

\textit{Conclusion}|We have introduced a generally applicable quasiprobability measure that interprets a Liouvillian eigenmode as a distribution of coherences. This extends the concept of biorthogonal quasiprobabilities, previously explored for non-Hermitian Hamiltonians \cite{engdahl1988complex}, to Liouville space. We have shown that dissipation does not necessarily erase semiclassical quantization. For an isotropic damped oscillator, each mode selects quantized orbits that are broadened in the presence of thermal noise, with measurable dynamical signatures. Additionally, quadratic open systems support dissipative analogs of invariant tori. As Lindblad dynamics is equivalent to a generalized Fokker-Planck evolution in phase space \cite{gardiner2004quantum, Carmichael1999, Strunz1998, Dubois2021, Wang2022}, our measure is also applicable to eigenstates of such a differential operator, which is of wider interest \cite{Risken1996, Gaspard1995, Chekroun2020, Eckmann2003, Herau2011, Edmunds2018, Davies2002}.

The coherence measure is averaged over all eigenmodes in the return amplitude; it would be interesting to explore whether the features of a given eigenmode|e.g., the peaks at nonzero $\varphi$ in Fig.~\ref{Fig4}(a)|can be isolated experimentally. Future studies can also examine the fate of the invariant tori under non-quadratic or integrability-breaking perturbations, which may lead to parallels of the Einstein-Brillouin-Keller (EBK) quantization \cite{Keller1985, Percival1977}. It would also be useful to understand the coherence structure when there is no classical limit, e.g., free fermions \cite{Prosen2008}, or when the classical limit features other attractors, e.g., a limit cycle \cite{Dutta2025} or a node with anisotropy \cite{Wolinsky1988}.


\begin{acknowledgments}
\vspace{1em}
{\it Acknowledgments}|We thank Arul Lakshminarayan and Bijay Agarwalla for stimulating discussions. FF acknowledges support by the European Union's Horizon Europe program under the Marie Sk{\l}odowska Curie Action GETQuantum (Grant No.~101146632). MH acknowledges support from the Deutsche Forschungsgemeinschaft under grant SFB 1143 (project-id 247310070). SD acknowledges travel support from the Anusandhan National Research Foundation (ANRF), India under the Prime Minister Early Career Research Grant (Project No.~ANRF/ECRG/2024/005069/PMS). AVT acknowledges support from the visiting students program at RRI.
\end{acknowledgments}

\vspace{1em}
{\it Data availability}|All figures can be reproduced directly from the presented closed-form expressions. The Mathematica notebook used to efficiently evaluate these expressions and make the plots is openly available \cite{zenodo}.

\bibliography{refs_v2.bib}

\twocolumngrid
\onecolumngrid
\vspace{1em}
\begin{center}
    \textbf{End Matter}
\end{center}
\vspace{1em}
\twocolumngrid

\appendix

\textit{Appendix A: Rate matrix and eigenvalues}|For a damped oscillator with Hamiltonian $\hat{H} = \omega \hat{a}^{\dagger} \hat{a} + \kappa \hat{a}^{\dagger} \hat{a}^{\dagger} \hat{a} \hat{a}$ and Lindblad operators $\hat{J}_1 = \sqrt{\gamma_1} \, \hat{a}, \hat{J}_2 = \sqrt{\gamma_2} \, \hat{a}^2$, the density matrix elements in the Fock basis are governed by a rate equation $\dot{\rho}^{(l)}_s = \sum_{s^{\prime}} \Gamma^{(l)}_{s,s^{\prime}} \rho^{(l)}_{s^{\prime}}$, where $\rho^{(l)}_s \coloneqq \rho_{s+l,s}$, $l,s \geq 0$, and the only nonzero transition rates are
\begin{align}
    \nonumber \Gamma^{(l)}_{s,s} = & -{\rm i}l [\omega + \kappa(2s+l-1)] - \gamma_1 (s+l/2) \\
    &- \gamma_2 [s(s-1) + (l/2)(2s+l-1)] \, , \\
    \Gamma^{(l)}_{s,s+1} = & \;\gamma_1 \sqrt{(s+1)(s+l+1)} \, , \\
    \Gamma^{(l)}_{s,s+2} = & \;\gamma_2 \sqrt{(s+1)(s+l+1)(s+2)(s+l+2)} \, .
\end{align}
As the rate matrix $\smash{\Gamma^{(l)}}$ is upper triangular, its eigenvalues are simply given by the diagonal entries \cite{Axler2024}, $\smash{\Lambda_{k,l} = \Gamma^{(l)}_{k,k}}$. The corresponding right eigenvector can be found using back substitution and has support over only the first $k+1$ elements. Conversely, the left eigenvector is found using forward substitution and has no support over the first $k$ elements. In the linear case of $\kappa = \gamma_2 = 0$, these have been worked out explicitly in Refs.~\cite{zubairy1980photon, Briegel1993, Endo2008, Honda2010}. In particular, the left eigenstate has the simple form $\hat{L}_{k,l} \propto (\hat{a}^{\dagger})^{k+l} \hat{a}^k$.

\vspace{1em}
\textit{Appendix B: Coherence matrix from normal modes}| The Liouvillian of a linearly damped harmonic oscillator is diagonalized in terms of the normal modes $a_K$ and $a_B$ via the similarity transformation $S = e^{-a_K a_B}$, as given in Eq.~\eqref{eq:normal_modes}. To see this, we use the Baker-Hausdorff lemma \cite{Sakurai2020} $e^A B e^{-A} = A + [A,B] + [A,[A,B]]/2! + \dots$ to obtain the pseudo-boson modes \cite{Trifonov2009, Bagarello2010, bagarello2017concise, Bagarello2022}
\begin{align}
    b_1 \coloneqq S a_K S^{-1} = a_K \, , \quad & b_1^{\prime} \coloneqq S a_K^{\dagger} S^{-1} = a_K^{\dagger} - a_B \, , \\
    b_2 \coloneqq S a_B S^{-1} = a_B \, , \quad\, & b_2^{\prime} \coloneqq S a_B^{\dagger} S^{-1} = a_B^{\dagger} - a_K \, .
\end{align}
Substituting these into Eq.~\eqref{eq:normal_modes} yields the expected Liouvillian. The eigenstates of $a_K$ and $a_B$ are simply dyads of Fock states of the original oscillator, since
\begin{align}
    a_K |n_K, n_B ) \leftrightarrow \hat{a} |n_K \rangle\langle n_B| \leftrightarrow \sqrt{n_K} \, |n_K-1, n_B ) \, , \\
    a_B |n_K, n_B ) \leftrightarrow |n_K \rangle\langle n_B| \hat{a}^{\dagger} \leftrightarrow \sqrt{n_B} \, |n_K-1, n_B ) \, .
\end{align}
The corresponding eigenstates of $\mathcal{L}$ are given by (up to the gauge freedom)
\begin{align}
    \nonumber & |\hat{R}_{n_K,n_B}) = S |n_K, n_B) = \textstyle \sum_{q=0}^{\infty} \frac{(-1)^q}{q!} a_K^q a_B^q |n_K, n_B) \\
    &\! = \sum_{q=0}^{\min(n_K, n_B)} \frac{(-1)^q}{q!} (n_K)_{\underline{q}} \, (n_B)_{\underline{q}} \, |n_K-q, n_B-q) \, , 
    \label{eq:damped_right} \\[0.5em]
    \nonumber & |\hat{L}_{n_K,n_B}) = (S^{-1})^{\dagger} |n_K, n_B) = \textstyle \sum_{q=0}^{\infty} \frac{1}{q!} \, a_K^{\dagger q} \, a_B^{\dagger q} \, |n_K, n_B) \\
    &\! = \sum_{q=0}^{\infty} \frac{1}{q!} (n_K)^{\underline{q}} \, (n_B)^{\underline{q}} \, |n_K+q, n_B+q) \, ,
\end{align}
where $\smash{(n)_{\underline{q}} \coloneqq \sqrt{n!/(n-q)!}}$ and $\smash{(n)^{\underline{q}} \coloneqq \sqrt{(n+q)!/n!}}$. The only dyad common to both eigenstates is $|n_K, n_B)$, so the coherence matrix in Eq.~\eqref{eq:coherence_matrix} simplifies to
\begin{equation}
    C_{n_K,n_B}(i,j) = ( i,j | \hat{R}_{n_K, n_B} )( i,j | \hat{L}_{n_K, n_B} )^* = \delta_{i,n_K} \delta_{j,n_B} \, .
\end{equation}

\vspace{1em}
\textit{Appendix C: Thermal distribution}|As discussed in the main text, for a thermal bath $\mathcal{L}$ has the same diagonal form with $S = e^{-a_K a_B} \smash{e^{n_T a_K^{\dagger} a_B^{\dagger}}}$ where $n_T \coloneqq 1/(e^{\beta\omega}-1)$, corresponding to the modified pseudo-boson modes
\begin{align}
    b_1 = (1+n_T) a_K - n_T a_B^{\dagger} \, , \quad & b_1^{\prime} = a_K^{\dagger} - a_B \, , \\
    b_2 = (1+n_T) a_B - n_T a_K^{\dagger} \, , \quad & b_2^{\prime} = a_B^{\dagger} - a_K \, .
\end{align}
As $S$ preserves the number difference $l = n_K - n_B$, both $\hat{R}_{n_K, n_B}$ and $\hat{L}_{n_K, n_B}$ live on the $l$-th diagonal in the Fock basis. For $l \geq 0$, $\smash{\hat{R}}$ has the matrix elements
\begin{align}
    & \nonumber ( j+l, j| \hat{R}_{n+l,n} ) \\
    &\! \nonumber = \sum_{p,q=0}^{\infty} \frac{(-1)^q}{q!} \frac{n_T^p}{p!} \,( j+l, j \big| (a_K a_B)^q \big(a_K^{\dagger} a_B^{\dagger}\big)^p \big| n+l, n ) \\
    &\! \nonumber = \sum_{p,q=0}^{\infty} \frac{(-1)^q}{q!} \frac{n_T^p}{p!} \frac{(n+p)! \, (n+p+l)!}{\sqrt{(n+l)! \, n! \, (j+l)! \, j!}} \, \delta_{n+p,j+q} \\
    &\! = \frac{(-1)^n n_T^j \sqrt{(n+l)! \, (j+l)!}}{(1+n_T)^{n+j+l+1} \; l! \sqrt{n! \, j!}} \, {}_2F_1(-n, -j; l+1; -1/n_T) \, ,
    \label{eq:thermal_right}
\end{align}
where ${}_2F_1$ is Gauss' hypergeometric function. The last line requires the use of linear transformation identities of ${}_2F_1$ \cite{DLMF}. Similarly, $\hat{L}$ has the matrix elements
\begin{align}
    & \nonumber ( j+l, j| \hat{L}_{n+l,n} ) \\
    &\! = (-n_T)^n \frac{\sqrt{(n+l)! \, (j+l)!}}{l! \sqrt{n! \, j!}} \, {}_2F_1(-n, -j; l+1; -1/n_T) \, .
    \label{eq:thermal_left}
\end{align}
One can also write these in terms of the Meixner polynomials \cite{Koekoek2010} $M_n(x; \mu, c)$ using the relation $M_n(x; \mu, c) = {}_2F_1(-n, -x; \mu; 1-1/c)$. Combining Eqs.~\eqref{eq:thermal_right} and \eqref{eq:thermal_left} yields the coherence distribution
\begin{align}
    & \nonumber C_{n+l,n}(j+l,j) \\
    & \!= \frac{e^{-\beta \omega (n+j)}}{(1\!+\!n_T)^{l+1}} \binom{n\!+\!l}{l} \binom{j\!+\!l}{l} \big[M_n \big(j; l\!+\!1, e^{-\beta \omega} \big) \big]^2 . 
\end{align}
Note this is symmetric under exchange of $n$ and $j$. Using orthogonality and recurrence relations of the Meixner polynomials \cite{Koekoek2010} one can find the mean $\overline{j}$ and standard derivation $\sigma_j$ as
\begin{equation}
    \overline{j} = \frac{n}{\tanh(\beta\omega/2)} + n_T (l+1) \, , \;\; 
    \sigma_j = \frac{\sqrt{2n^2 \!+\! (2n \!+\! 1)(l \!+\! 1)}}{2 \sinh(\beta \omega / 2)} \, .
\end{equation}
Furthermore, the ``turning points'' $j_{\pm}$ in Fig.~\ref{Fig3}(a) can be estimated from large-$n$ asymptotics \cite{Jin1997},
\begin{equation}
    j_- \approx n \tanh(\beta \omega / 4) \, , \quad j_+ \approx n \coth(\beta \omega / 4) \, .
\end{equation}

\begin{figure}
    \includegraphics[width=1\columnwidth]{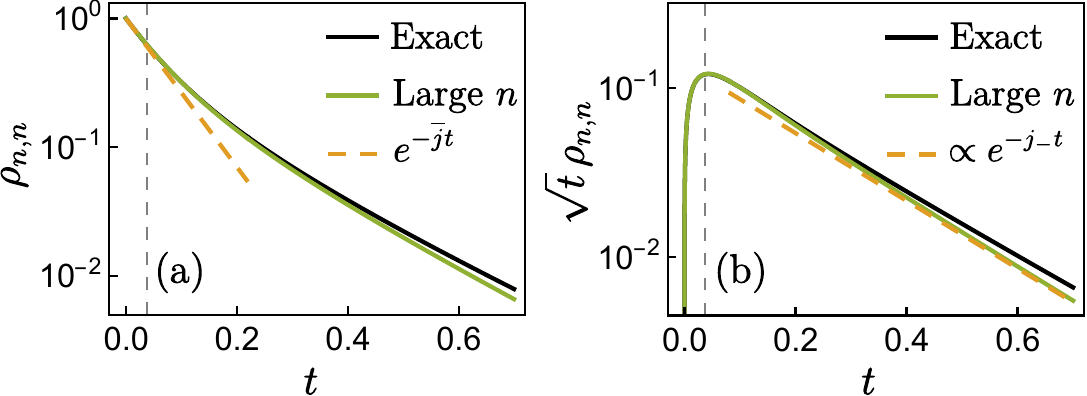}
    \caption{Return probability of $|n=10\rangle$ for $\beta \omega = 2, \gamma=1$, comparing exact [Eq.~\eqref{eq:return_exact}] and large-$n$ [Eq.~\eqref{eq:return_large_n}] predictions. The gray line at $t = 1/(2 \overline{j})$ is where $\sqrt{t} \rho_{n,n}$ peaks for $\beta \omega \gtrsim 1$.}
    \label{Fig5}
\end{figure}

The return probability of a Fock state $|n\rangle$ is given by $\rho_{n,n}(t) = \sum_j C_{j,j}(n,n) e^{-j \gamma t}$. Using Weisner's bilinear generating function for hypergeometric polynomials \cite{Weisner1955},
\begin{equation}
    \rho_{n,n}(t) \propto \frac{(e^{-\gamma t} - e^{-\beta \omega})^n}{(1-e^{-\gamma t - \beta\omega})^{n+1}} 
    P_n \Big(\frac{\cosh \beta\omega + \cosh \gamma t - 2}{\cosh \beta\omega - \cosh \gamma t} \Big) ,
    \label{eq:return_exact}
\end{equation}
where the proportionality constant is $1-e^{-\beta\omega}$ and $P_n$ is a Legendre polynomial. For $n \gg 1$ the decay is dominated by the bulk of the distribution between $j_-$ and $j_+$, where $C_{n,n}(j,j)$ has an arcsine envelope \cite{Wang2009}, yielding
\begin{equation}
    \rho_{n,n}(t) \approx \! \int_{j_-}^{j_+} \!\! \frac{e^{-j \gamma t} \, {\rm d}j}{\pi \sqrt{(j \!-\! j_-)(j_+ \!-\! j)}} 
    = e^{-\overline{j} \gamma t} I_0(\gamma_d t) \, ,
    \label{eq:return_large_n}
\end{equation}
where $\gamma_d \coloneqq (j_+ - j_-) \gamma /2$ and $I_0$ is a modified Bessel function of the first kind. From the properties of $I_0$ \cite{DLMF}, $\rho_{n,n} \approx \smash{e^{-\overline{j} \gamma t}}$ for $\gamma_d t \ll 1$ and $\rho_{n,n} \approx \smash{e^{-j_- \gamma t}} / \sqrt{2 \pi \gamma_d t}$ for $\gamma_d t \gg 1$. Thus, $\sqrt{t} \rho_{n,n}(t)$ exhibits a peak at $\gamma t = O(1/n)$ followed by an exponential decay at a rate $j_- \gamma$ (Fig.~\ref{Fig5}). Additionally, as shown in Fig.~\ref{Fig3}(b), the instantaneous decay rate of $\rho_{n,n}$ exhibits a shoulder close to $j_-$.

\vspace{1em}
\textit{Appendix D: Properties of the super Husimi function}| We have defined the super Husimi function for an eigenmode $p$ as $C_p(\alpha_K, \alpha_B) \coloneqq ( \alpha_K, \alpha_B | \mathcal{F}_p | \alpha_K, \alpha_B )$. Inserting completeness of the Fock states gives
\begin{align}
    \nonumber & C_p(\alpha_K, \alpha_B) \\
    \nonumber & \!\!= \sum_{i, j, i^{\prime}, j^{\prime}} ( \alpha_K, \alpha_B | i, j )( i,j | \mathcal{F}_p | i^{\prime}, j^{\prime} )( i^{\prime}, j^{\prime} | \alpha_K, \alpha_B ) \\
    & \!\!= e^{-|\alpha_K|^2 -|\alpha_B|^2} \!\! \sum_{i, j, i^{\prime}, j^{\prime}} \! ( i,j | \mathcal{F}_p | i^{\prime}, j^{\prime} ) \frac{(\alpha_K^*)^i \alpha_B^j \alpha_K^{i^{\prime}} (\alpha_B^*)^{j^{\prime}}}{\sqrt{i! \, j! \, i^{\prime}! \, j^{\prime}!}} \, .
    \label{eq:husimi_expansion}
\end{align}
Two results follow from this expansion. First, with $\alpha_K = r_K e^{{\rm i} \phi_K}$ and $\alpha_B = r_B e^{{\rm i} \phi_B}$, the radial marginal reduces to
\begin{align}
    \nonumber & A_p(r_K, r_B) \\
    & \!\! \coloneqq \!\int\!\! {\rm d}\phi_K \!\!\int\!\! {\rm d}\phi_B \, C_p(\alpha_K, \alpha_B) = \sum_{i,j} C_p(i,j) Q_i(r_K) Q_j(r_B) \, ,
\end{align}
where $Q_n(r) \coloneqq 2 \pi e^{-r^2} r^{2n} / n!$. Second, if the dynamics is isotropic, the right and left eigenstates belong to the same diagonal in Fock space, i.e., $i-j = i^{\prime}-j^{\prime}$ in Eq.~\eqref{eq:husimi_expansion}, so the super Husimi function depends only on the relative phase $\varphi = \phi_K - \phi_B$, not on $\phi_K$ and $\phi_B$ separately.

For a linearly damped harmonic oscillator, $C_p(\alpha_K, \alpha_B)$ can be found in closed form. From Eq.~\eqref{eq:damped_right},
\begin{align}
    \nonumber & ( \alpha_K, \alpha_B | \hat{R}_{n_K, n_B} ) \\
    & \!= \sqrt{n_K! \, n_B!} \, e^{-\frac{|\alpha_K|^2 + |\alpha_B|^2}{2}} \sum_{q=0}^{n_{\min}} \frac{(-1)^q (\alpha_K^*)^{n_K-q} \alpha_B^{n_B-q}}{q! \, (n_K-q)! \, (n_B-q)!} \, ,
\end{align}
where $n_{\min} \coloneqq \min(n_K, n_B)$. Conversely, using the fact that $|\alpha_K, \alpha_B)$ is a joint eigenstate of $a_K$ and $a_B$,
\begin{align}
    \nonumber & ( \hat{L}_{n_K, n_B} | \alpha_K, \alpha_B ) = ( n_K, n_B | e^{a_K a_B} | \alpha_K, \alpha_B ) \\
    & \!= e^{\alpha_K \alpha_B^* - \frac{|\alpha_K|^2 + |\alpha_B|^2}{2}} \alpha_K^{n_K} (\alpha_B^*)^{n_B} / \sqrt{n_K! \, n_B!} \, .
\end{align}
Hence, the super Husimi function is given by
\begin{align}
    \nonumber & C_{n_K, n_B}(\alpha_K, \alpha_B) \\
    & \!= e^{r_K r_B e^{{\rm i}\varphi} -r_K^2 - r_B^2} 
    \sum_{q=0}^{n_{\min}} \frac{(-1)^q \, r_K^{2n_K-q} \, r_B^{2n_B-q} \, e^{{\rm i} q \varphi}}{q! \, (n_K-q)! \, (n_B-q)!} \, .
    \label{eq:super_husimi_sum}
\end{align}
For $n_K \geq n_B$ it can be written more compactly as
\begin{equation}
    C_{n+l, n}(\alpha_K, \alpha_B) = \frac{e^{\alpha_K \alpha_B^*} |\alpha_K|^{2l} (-\alpha_K \alpha_B^*)^n}{e^{|\alpha_K|^2 + |\alpha_B|^2} (n+l)!}  L_n^{(l)}(\alpha_K^* \alpha_B) \, ,
    \label{eq:super_husimi}
\end{equation}
where $\smash{L_n^{(l)}}$ is a generalized Laguerre polynomial. For large $n$ the peaks of this distribution can be found using asymptotic expansions of $\smash{L_n^{(l)}}$ \cite{Qiu2008}. For $l=0$ they occur at $r_K = r_B = \sqrt{n \rho}$ where $\rho = O(1)$. From Eq.~\eqref{eq:super_husimi},
\begin{equation}
    \frac{1}{n} \ln |C_{n,n}| = \rho (\cos\varphi-2) + \ln \rho + \frac{1}{n} \ln \!\big| L_n^{(0)} \!\big( n \rho e^{-{\rm i} \varphi} \big) \!\big|
\end{equation}
up to a constant. Using the asymptotic form \cite{Qiu2008}
\begin{equation}
    \lim_{n \to \infty} \frac{1}{n} \ln \big| L_n^{(0)} \big( 2 n (1+\cosh z) \big) \big| = \text{Re} (1+z+e^{-z}) 
\end{equation}
$\forall \;\text{Re}(z) > 0$ gives the peaks at $\rho^{\star} = 1+1/\sqrt{2} \approx 1.71$ and $\varphi^{\star} = \pm \cos^{-1}(\sqrt{2}-1/2) \approx \pm 0.13\pi$, as shown in Fig.~\ref{Fig4}(a).

The return amplitude of a dyad $|\alpha_K \rangle \langle \alpha_B|$ is given by $\zeta(t) = \sum_{n_K, n_B} C_{n_K, n_B} (\alpha_K, \alpha_B) \, e^{\Lambda_{n_K, n_B} t}$. Substituting the eigenvalues $\Lambda_{n_K, n_B}$ from Eq.~\eqref{eq:normal_modes}, one can write
\begin{equation}
    \zeta(t) = G \big( e^{-(\gamma/2 + {\rm i} \omega)t}, e^{-(\gamma/2 - {\rm i} \omega)t} \big) \, ,
    \label{eq:return_amplitude}
\end{equation}
where $G(u,v) \coloneqq \sum_{n_K, n_B} C_{n_K, n_B} u^{n_K} v^{n_B}$ is the generating function. The dependence on $\alpha_K$ and $\alpha_B$ is implicit. Using the expression for $C_{n_K, n_B}$ from Eq.~\eqref{eq:super_husimi_sum} gives
\begin{equation}
    G(u,v) = e^{(1-uv) \alpha_K \alpha_B^* - (1-u)|\alpha_K|^2 - (1-v)|\alpha_B|^2} \, .
    \label{eq:generating_function}
\end{equation}
From Eqs.~\eqref{eq:return_amplitude} and \eqref{eq:generating_function}, we find $\dot{\zeta}(0) = -(\frac{\gamma}{2} + {\rm i}\omega) \overline{n_K} - (\frac{\gamma}{2} - {\rm i}\omega) \overline{n_B}$, where $\overline{n_K}$ and $\overline{n_B}$ are the first moments,
\begin{align}
    \overline{n_K} = \partial_u G(u,v)|_{u=v=1} = |\alpha_K|^2 - \alpha_K \alpha_B^* \, , \\
    \overline{n_B} = \partial_v G(u,v)|_{u=v=1} = |\alpha_B|^2 - \alpha_K \alpha_B^* \, .
\end{align}
Hence, the initial decay rate of the Loschmidt echo is
\begin{equation}
    -\partial_t \ln |\zeta|^2 |_{t=0} = - 2 \,\text{Re}\, \dot{\zeta}(0) = \gamma |\alpha_K - \alpha_B|^2 \, .
\end{equation}
Although this rate and the first moments vanish for $\alpha_K = \alpha_B = r e^{{\rm i}\theta}$, the distribution is not peaked at $n_K = n_B = 0$. Instead, from Eq.~\eqref{eq:super_husimi}, it is highly oscillatory with a magnitude $|C_{n+l, n}| \propto u^{n+l} |\smash{L_n^{(l)}(u)}| / (n+l)!$, where $u \coloneqq r^2$. For $u < 4n+2l+2$, the Laguerre polynomial oscillates with the envelope \cite{DLMF} $\smash{E_n^{(l)}(u)} = e^{u/2} (n/u)^{l/2} (\pi^2 nu)^{-1/4}$, which gives a peak of $|C_{n+l, n}|$ at $l^{\star}=0, n^{\star} \approx u$ for $u \gg 1$, as seen in Fig.~\ref{Fig4}(b). On the other hand, for $\varphi=0$ the distribution is positive semidefinite [see Eq.~\eqref{eq:super_husimi_sum}], so the peaks coincide with the first moments at $n_K^{\star} \approx r_K^2 + r_K r_B$ and $n_B^{\star} \approx r_B^2 + r_K r_B$ for $u \gg 1$ [Fig.~\ref{Fig4}(c)].

\end{document}